\documentclass[letterpaper, 10 pt, conference]{ieeeconf}  

\IEEEoverridecommandlockouts                              

\usepackage{amsmath} 
\usepackage{amssymb}  
\usepackage{cite}
\usepackage{graphicx}
\usepackage[caption=false, font=footnotesize]{subfig}
\usepackage{tikz}
\usepackage{pgfplots}

\newcommand\copyrighttext{%
	\footnotesize Copyright $\copyright$ 2019 IEEE.
	Personal use of this material is permitted.
	Permission from IEEE must be obtained for all other uses, in any current or future media, including reprinting/republishing this material for advertising or promotional purposes, creating new collective works, for resale or redistribution to servers or lists, or reuse of any copyrighted component of this work in other works. }%
\newcommand\copyrightnotice{%
	\begin{tikzpicture}[remember picture,overlay]%
	\node[anchor=south,yshift=10pt] at (current page.south) {\fbox{\parbox{\dimexpr\textwidth-2cm}{\copyrighttext}}};%
	\end{tikzpicture}%
	\vspace{-10pt}%
}

\title{\LARGE \bf
A Trust Management and Misbehaviour Detection Mechanism for Multi-Agent Systems and its Application to Intelligent Transportation Systems$^\ast$
}

\author{Johannes M{\"u}ller$^{1}$, Tobias Meuser$^{2}$, Ralf Steinmetz$^{2}$, and Michael Buchholz$^{1}$
\thanks{*Part of this work was financially supported by the Federal Ministry of Economic Affairs and Energy of Germany within the program "Highly and Fully Automated Driving in Demanding Driving Situations" (project MEC-View, grant number 19A16010I).}	
\thanks{$^{1}$Johannes M{\"u}ller and Michael Buchholz are with Institute of Measurement, Control and Microtechnology,
	Ulm University, D-89081 Ulm, Germany
	{\tt\small \{johannes-christian.mueller, michael.buchholz\}@uni-ulm.de}
}%
\thanks{$^{2}$Tobias Meuser and Ralf Steinmetz are with the Multimedia Communications Lab (KOM), Technische Universit{\"a}t Darmstadt, D-64283 Darmstadt, Germany
	{\tt\small \{tobias.meuser, ralf.steinmetz\}@kom.tu-darmstadt.de}}%
}

\begin{document}

\maketitle
\copyrightnotice%
\thispagestyle{empty}
\pagestyle{empty}

\begin{abstract}

Cooperative information shared among a multi-agent system (MAS) can be useful to agents to efficiently fulfill their missions. Relying on wrong information, however, can have severe consequences. While classical approaches only consider measurement uncertainty, reliability information on the incoming data can be useful for decision making. In this work, a subjective logic based mechanism is proposed that amends reliability information to the data shared among the MAS. 
If multiple agents report the same event, their information is fused. In order to maintain high reliability, the mechanism detects and isolates misbehaving agents. Therefore, an attacker model is specified that includes faulty as well as malicious agents. The mechanism is applied to Intelligent Transportation Systems (ITS) and it is shown in simulation that the approach scales well with the size of the MAS and that it is able to efficiently detected and isolated misbehaving agents. 
\end{abstract}


\section{INTRODUCTION}

Consider, for example, an intersection scenario where an agent, in this case an intelligent vehicle (IV), can use cooperative information, provided by an other agent, e.g. a Road Side Unit (RSU), to smoothly merge into an occluded traffic gap. If the information received from the RSU is correct, the IV can use this information to increase its efficiency. However, if it relies on faulty information, the IV will assume a traffic gap where there is none and hence need to do an emergency braking. In the worst case the car will crash while merging.

The example illustrates first that the cooperative information must be highly reliable to operate upon it, and second that measurement uncertainty, often probabilistically modeled in terms of means and covariances is not sufficient to express reliability: in case of the intersection scenario, a very precise, low variance description of an approaching traffic gap is worthless if a faulty sensor in the RSU oversees a further vehicle within that traffic gap. 

Subjective logic \cite{Joesang2016} is a framework that extends classical probability by an uncertainty measure reflecting all the shortcomings mentioned above, dealing with so-called opinions. A brief summary about the theoretical background on subjective logic is given in Section \ref{SL_basics}. The reliability of a piece of information then results from the uncertainty the information source itself poses on its information, the trust in the information source, and the amount of evidence supporting the piece of information.

In MAS, trust is usually established by reputation systems \cite{Yu2013}. However, as common reputation systems are always based on previous experience, they are incapable to quickly react to a sudden change of reliability, which might be required due to a sensor fault or a cyber attacker taking over an agent. Hence, common reputation systems are not sufficient to guarantee the reliability of incoming cooperative information.
By combining a classical reputation system with a Trust Revision Algorithm similar to that described in \cite{Joesang2017}, the mechanism proposed in this work is capable to react quickly to a change in the reliability of a node. While \cite{Joesang2017} proposes to do an averaging belief fusion over all available opinions to achieve the reference opinion used to judge the trustworthiness of the opinions, we propose to do a clustering of all available opinions first and then calculate the reference opinion by applying average belief fusion only to the cluster with the most opinions. Hence, a majority of low trust agents can challenge even a high reputation misbehaving node. Furthermore, to account for aging of information, we incorporate the mechanism described in \cite{Meuser2019}. The details of the algorithm are described in Section \ref{Algorithm}.

In order to maintain a high level of reliability, a mechanism is needed that detects and isolates misbehaving agents. In the context of MAS control, such mechanisms are referred to as Fault Detection and Isolation (FDI) Methods \cite{Zidan2017}. However, intentionally malicious agents have not been included so far. Furthermore, according to the MAS survey \cite{Dorri2018} published in October 2018, FDI methods still suffer from the following severe limitations: 1) most mechanisms assume homogeneous agents, while in an open MAS such as ITS, this assumption is not justified, 2)  exhaustive resource consumption, 3) isolating faulty nodes is an open issue and 4) most of the proposed solutions are centralized by nature. The mechanism described within this work offers a proposition how these shortcomings can be addressed.

In the field of communications, some research efforts have been dedicated to misbehavior detection including security issues such as privacy of the users and intrusion detection of cyber attacks. In his recent survey \cite{Heijden2018}, published in October 2018, van der Heijden gives a broad and detailed overview on current state-of-the-art misbehavior detection mechanisms. The survey defines misbehaving agents as nodes sending wrong information through the MAS, either due to a sensor fault, or because they are malicious, i.e. cyber attackers. However, it is stated in the survey that no ideal solution to misbehavior detection has been published so far. A lot of approaches also suffer badly from their exhaustive resource consumption \cite{Leinmueller2010}, \cite{Rawat2011}. Furthermore, the threshold when to report a malicious message, the usage of pseudonyms, and the traceability of users by attackers are mentioned as open research questions. Within this work, possible answers to these questions are provided. Furthermore, it is shown in large scale simulation that the approach scales well and is efficient in terms of resource usage.

In this work, we settle on the idea of Dietzel et al. \cite{Dietzel2014} to use subjective logic for misbehavior detection. However, in contrast to building up a very general, but possibly resource inefficient framework, the mechanism presented in this work proposes a trade-off between security, generality, and resource consumption. The system setup, the attacker model, and the trade-off considerations are described in Section \ref{SystemOverview}.

The key contributions of this paper can be summarized as following:
\begin{itemize}
	\item A trust management and misbehavior mechanism is presented which overcomes the shortcomings described in  \cite{Yu2013} and \cite{Dorri2018}. This is, our proposition 1) is able to cope with heterogeneous agents, 2) shows little resource usage, 3) can isolate faulty agents, and 4) quickly reacts to sudden changes of reliability.
	\item Propositions to the open research questions 1) when to report misbehavior, 2) how to deal with pseudonyms, and 3) how to prevent attackers from tracing other agents, as formulated in \cite{Heijden2018}, are given.
	\item The proposed trust management and misbehavior detection mechanism is able to provide reliability information on all data sent through the MAS.
\end{itemize}


\section{SUBJECTIVE LOGIC BASICS} \label{SL_basics}

In this section, we briefly summarize the subjective logic basics used in this paper. The definitions and theorems are taken from \cite{Joesang2016}, where further details can be found.
%
%

\textbf{Definition 1 Subjective logic opinion:} Let $\mathbb{X}$ be a domain and $\text{card}\{\mathbb{X}\} \geq 2$, where $\text{card}\{ \, \cdot \, \}$ is the cardinality. Let $X$ further be a random variable in $\mathbb{X}$. A subjective logic opinion (opinion in short) is an ordered triple $\omega_x = (\mathbf{b}_x,u_x,\mathbf{a}_x)$ with
\begin{subequations}
	\begin{align}
		\mathbf{a}_x : \mathbb{X} \mapsto [0,1], \qquad 1 &= \sum\limits_{x \in \mathbb{X}} \mathbf{a}_x \, ,\\
		\mathbf{b}_x : \mathbb{X} \mapsto [0,1], \qquad 1 &= u_x + \sum\limits_{x \in \mathbb{X}} \mathbf{b}_x \, .
	\end{align}
\end{subequations}
Hereby, $\mathbf{b}_x$ is the belief mass distribution over $\mathbb{X}$, $u_x$ is the uncertainty mass representing a lack of evidence and $\mathbf{a}_x$ is the base rate distribution over $\mathbb{X}$ representing the prior. 
%
%

\textbf{Definition 2 Dirichlet Distribution:} Let $\mathbb{X}$ be a domain of $W$ mutually disjoint values, $\mathbf{r}_x \in \mathbb{N}$ be the evidence for outcome $x \in \mathbb{X}$, $\mathbf{a}_x$ a prior distribution over $\mathbb{X}$, and $\mathbf{p}_x$ the probability distribution of $x$ over $\mathbb{X}$. Then, the probability density function (PDF)
\begin{equation}
	\text{Dir}(\mathbf{p}_x,\mathbf{r}_x,\mathbf{a}_x) = \frac{\Gamma \left( \sum \limits_{x \in \mathbb{X}} (\mathbf{r}_x + \mathbf{a}_x W) \right)}{\prod \limits_{x \in \mathbb{X}} \Gamma ( \mathbf{r}_x + \mathbf{a}_x W ) } \prod \limits_{x \in \mathbb{X} } \mathbf{p}_x ^{\mathbf{r}_x + \mathbf{a}_x W - 1} \, ,
	\label{eq:Dirichlet}
\end{equation} 
where $\mathbf{r}_x + \mathbf{a}_x W \geq 0$ and $\mathbf{p}_x > 0$ for $\mathbf{r}_x + \mathbf{a}_x W < 1$, is called Dirichlet PDF. A Dirichlet PDF with $W=2$ is called $\beta$-distribution. In (\ref{eq:Dirichlet}), $\Gamma( \, \cdot \, )$ is the well known Gamma function \cite{Bronshtein2007}.
%
%

\textbf{Definition 3 Aleatory Cumulative Belief Fusion:} Let $\omega_x^A$ and $\omega_x^B$ be source A and B's respective opinions over the same variable $X$ on domain $\mathbb{X}$. Let $\omega_x^{A \diamond B}$ be the opinion such that
\begin{equation}
\omega_x^{A \diamond B} = \left \{  \begin{array}{ll}
\mathbf{b}_x &= \frac{ \mathbf{b}_x^{A} u_x^{B} + \mathbf{b}_x^{B} u_x^{A} }{ u_x^{A} + u_x^{B} - u_x^{A} u_x^{B} } \\
\\
u_x &= \frac{ u_x^{A} u_x^{B} }{ u_x^{A} + u_x^{B} - u_x^{A} u_x^{B} }\\
\\
\mathbf{a}_x &= \frac{ \mathbf{a}_x^{A} u_x^{B} + \mathbf{a}_x^{B} u_x^{A} - (\mathbf{a}_x^{A} + \mathbf{a}_x^{B}) u_x^{A} u_x^{B} }{ u_x^{A} + u_x^{B} - 2 u_x^{A} u_x^{B} } 	
\end{array}\right . \, ,
\end{equation}
where $0 < u_x^A < 1$ and $0 < u_x^B < 1$, then the operator $\oplus$ in
\begin{equation}
\omega_x^{A \diamond B} = \omega_x^A \oplus \omega_x^B
\end{equation}
is called Aleatory Cumulative Belief Fusion.
%
%

\textbf{Definition 4 Aleatory Average Belief Fusion:} Let $\omega_x^A$ and $\omega_x^B$ be source A and B's respective opinions over the same variable $X$ on domain $\mathbb{X}$. Let $\omega_x^{A \diamond B}$ be the opinion such that
\begin{equation}
\omega_x^{A \diamond B} = \left \{  \begin{array}{ll}
\mathbf{b}_x &= \frac{ \mathbf{b}_x^{A} u_x^{B} + \mathbf{b}_x^{B} u_x^{A} }{  u_x^{A} \, + \, u_x^{B} }  \\
\\
u_x &= \frac{2 u_x^{A} u_x^{B} }{ u_x^{A} \, + \, u_x^{B} }  \\
\\
\mathbf{a}_x &= \frac{ \mathbf{a}_x^{A} \, + \, \mathbf{a}_x^{B} }{ 2 } 	
\end{array}\right . \, ,
\end{equation}
where $u_x^A \neq 0$ and $u_x^B \neq 0$, then the operator $\underline{\oplus}$ in
\begin{equation}
\omega_x^{A \diamond B} = \omega_x^A \, \underline{\oplus} \, \omega_x^B
\end{equation}
is called Aleatory Average Belief Fusion. When multiple opinions are fused, we use the shorthand 
\begin{equation}
	\overset{N}{\underset{i=1} {\underline{\bigoplus}} } \, \omega_x^{A_i} = \omega_x^{A_1} \, \underline{\oplus} \, \omega_x^{A_2} \underline{\oplus} \dots \underline{\oplus}  \omega_x^{A_N} \, .
\end{equation}

\textbf{Theorem 1 Equivalent mapping:} Let $\omega_x = (\mathbf{b}_x,u_x,\mathbf{a}_x)$ be a subjective logic opinion and $\text{Dir}(\mathbf{p}_x,\mathbf{r}_x,\mathbf{a}_x)$ a Dirichlet PDF over the same $x \in \mathbb{X}$. Then, the equivalent mapping
\begin{equation}
	\small
	\label{eq:Mapping}
	\left . \begin{matrix}
	\mathbf{b}_x = \frac{\mathbf{r}_x}{W + \sum \limits_{x_i \in \mathbb{X}} \mathbf{r}_x} \\
	\\
	u_x = \frac{W}{W + \sum \limits_{x_i \in \mathbb{X}} \mathbf{r}_x} 
	\end{matrix} \right \} \iff
	\left \{ \begin{array}{ll}
	\mathbf{r}_x &= \frac{ W \mathbf{b}_x}{u_x} \\
	\\
	1 &= u_x + \sum \limits_{x_i \in \mathbb{X}} \mathbf{b}_x(x_i)
	\end{array} \right . \, ,	
\end{equation}
transforms the Dirichlet distribution into the subjective logic opinion and vice versa.
%

\textbf{Theorem 2 Probability projection:} Let $\omega_x = (\mathbf{b}_x,u_x,\mathbf{a}_x)$ be a subjective logic opinion and $\text{Dir}(\mathbf{p}_x,\mathbf{r}_x,\mathbf{a}_x)$ a Dirichlet PDF over the same $x \in \mathbb{X}$. Then, the projection
\begin{equation}
	\mathbf{P}_x = \mathbf{b}_x + \mathbf{a}_x u_x
\end{equation}  
is optimal with respect to the maximum likelihood operator on $\text{Dir}(\mathbf{p}_x,\mathbf{r}_x,\mathbf{a}_x)$.
\section{SYSTEM OVERVIEW} \label{SystemOverview}


 For this work, we assume a cellular network, as cellular network topologies are frequently used to realize connected intelligent transportation systems (ITS) \cite{Heijden2018}. In order to reduce bandwidth consumption, a pushed based communication scheme is used. This means that the providers of cooperative information can publish their observations to a topic the users can subscribe to. A broker  routes the information from providers to users. In the context of ITS,  a topic can be a map segment, e.g. of an intersection or a highway lane. The topic association is assumed to be stored in the digital map of the IVs. Additionally, providers/users have to specify, for which period of time they can offer/need cooperative information on the topic they publish/subscribed to. Note that the broker acts as an abstraction layer between providers and users. Hence, neither users nor providers know which other agents they are interacting with. By using a public key infrastructure (PKI), as specified by ETSI TS 103 097 standard \cite{ETSI} or IEEE 1609.2 standard \cite{IEEE1609}, message authenticity is guaranteed. In order to maintain the agents' privacy, users and providers may have multiple, frequently changing pseudonyme identities, with which they are connecting to the broker. The broker, however, holds a direct mapping between the pseudonyms and the true, longterm identities. Hence, by resolving the true identities of each agent publishing or subscribing to a topic, so-called Sybil attacks (see Section \ref{AttackerModel}) can be easily detected.

Basically, there are three groups of agents present in the MAS: the providers sending cooperative information, the users fusing incoming cooperative information and  operating upon them, and the broker distributing the cooperative information through the network. In case of data inconsistency, the receiving user reports this to the broker. The broker then decides, in a court-case like procedure as neutral judge according to the available evidence, which agent is misbehaving. The trust into the misbehaving agent then is revised. In severe cases, the broker revokes the agent's certificate leading to a ban of the respective agent. 

In order to be able to resolve word-against-word situations, where the same number of agents state that the other part has shown misbehavior, the broker may keep some hidden observers. Hidden observers are specially trusted agents that only listen to the messages of other agents and measure their environment without sending their information. However, if a misbehavior is detected, the hidden observers report the incidence. This way, harm can be avoided before an attack is successful and attackers can never be sure whether there is a further information source disclosing their attack.

The communication procedure can be summarized as follows:
\begin{enumerate}
	\item At least one provider of cooperative information publishs to a certain topic specifying a time interval during which it can provide information.
	\item At least one user subscribes to a certain topic specifying in which time interval it needs cooperative information.
	\item The broker resolves the pseudonyms and verifies that each agent is only registered with one pseudonym. Else, the respective misbehaving agent is classified as misbehaving.
	\item If there are providers available that offer cooperative information for a time interval matching the needs of users, the broker routes the respective cooperative information from the providers to the users.
	\item The users fuse the incoming cooperative information. If the data is consistent, the users act upon them. Else, the users report the incident to the broker. If an incident is reported, the broker temporarily revokes the certificates of all agents connected to the reported incidence. Then the broker itself runs the misbehavior detection mechanism to verify the claims and revise the trust into the involved agents accordingly
	\item After successfully completing the communication to acquire the 
	cooperative information, the user reports this to the broker. The broker then increases trust of all involved agents as described in Section \ref{TrustBuilding}. 
\end{enumerate}

\subsection{Attacker Model} \label{AttackerModel}

In order to demonstrate the effectiveness of the trust management and misbehavior detection mechanisms, it has to be stated what kinds of attacks are assumed. Until now, no universally accepted attacker model consistently used for cooperative ITS has been established so far \cite{Heijden2018}. Furthermore \cite{Heijden2018} also states that the classial Dolev-Yao attacker model is to strong. As the main focus of this work is the detection of faults rather than sophisticated cyber attacks, the following simple attacker model is assumed, still matching most of the state-of-the-art assumptions \cite{Heijden2018}:
\begin{itemize}
	\item Attackers are allowed to send semantically incorrect data while sticking to the protocol (e.g. sensor fault)
	\item Attackers might use several pseudonyms in parallel (Sybil attack).
	\item Attackers must be agents. Hence, although they might have extended computational power or increased transmission range, the attackers physical presence is limited.
	\item A local honest majority is assumed. We weaken this assumption by introducing hidden observers. As demonstrated in Section \ref{IntersectionExample}, thus even collaborative attacks, where two attackers coordinate their efforts, can be detected by another agent and an RSU.
	\item As a cellular network topology is chosen, the broker routing the cooperative information from providers to users is assumed to be honest. 
\end{itemize}
 
\subsection{Trade-off Considerations}

As stated in \cite{Heijden2018}, the ideal solution for misbehavior detection schemes has not been presented yet. Sophisticated attacker models require complex defense mechanisms that are very resource consuming. The assumption of a certain network topology allows for resource optimization towards that specific setting at cost of generality. 

Deciding for a cellular network topology reduces generality and, due to the assumption of a honest broker, also safety. However, from a practical perspective, a cellular based approach is commonly used and the assumption of a honest broker appears reasonable, because in practice the network operators would provide the broker. As network operators have far more sophisticated methods at hand, effectively protecting against a malicious network provider would need skyrocketing amounts of efforts anyway. In turn, it is easy for the legal authorities to check the correct behavior by frequent inspections. The network overhead is limited to the PKI overhead. This overhead, however, is unavoidable if message authenticity should be guaranteed \cite{Heijden2018}. A further benefit of push-based communication is that anonymity between agents other than the broker can be easily set up. As the broker knows the direct mapping between true identity and pseudonyms, the broker can easily detect Sybil attacks.
Hidden agents are expensive to operate. However, a small number of them is sufficient to increase the risk of the agents to be detected. Furthermore with the help of hidden observers, the assumption of a local honest majority can be softened. Thus, a small number of hidden observers is a good trade-off between operational cost and increased security.


\section{TRUST MANAGEMENT AND MISBEHAVIOR DETECTION MECHANISM} \label{Algorithm}

The trust management and misbehavior detection mechanism is used to revise the trust into agents that send conflicting information and to detect misbehavior. It basically consists of two parts: the trust building part and the trust revision part. Trust is build up whenever an agent has acted correctly, while trust into an agent is revised, when its reported opinion is inconsistent with the majority of other agents' opinions reporting on the same topic.

\subsection{Trust Building} \label{TrustBuilding}

Whenever multiple communicating agents reach a consensus opinion, all participating agents are rewarded with an increase of trust into them. Furthermore, whenever the broker decides that an agent has acted correctly during a reported misbehavior case, the agent is strongly rewarded. However, successfully completed cooperations between the same agents is not statistically independent. Furthermore, the trust into an agent should fade into uncertainty over time, because an agent that has proven to behave well long time ago due to sensor degradation might not be reliable in the presence any more. This behavior is modeled using the mapping between $\beta$-distributions and binomial opinions, Eq. (\ref{eq:Mapping}).

A $\beta$-distribution is parametrized by $r_x$ and $s_x$, where $s_x$ describes the recorded favorable outcomes of a statistical process, while $s_x$ describes the unfavorable outcomes. Hence, each successfully completed cooperation leads to an increment of $r_x$ by 1. If the broker finds that an agent has behaved correctly through a trust revision case, its trust is increased by $w_{\text{TR}} \in \mathbb{N}$. In order to account for the statistical dependence of successful outcomes between the same group of agents, the successfully completed cooperations are accumulated using the consensus operator for partially dependent $\beta$-PDFs \cite{Joesang2006} with dependence factor $\lambda^{A_k}$. This results in the splitting of an opinion into an independent part $r_x^{A_{k,\text{ind}}}$, $r_x^{A_{k,\text{ind}}}$ and a dependent part $r_x^{A_{k,\text{dep}}}$, $s_x^{A_{k,\text{dep}}}$ according to
\begin{equation}
		\label{eq:PartiallyDependendCBF}
		\left \{ \begin{matrix}
		r_x^{A_{k,\text{ind}}} = r_x^{A_k} (1 - \lambda^{A_k}) \\
		s_x^{A_{k,\text{ind}}} = s_x^{A_k} (1 - \lambda^{A_k})
		\end{matrix} \right . \, , \qquad 		\left \{ \begin{matrix}
		r_x^{A_{k,\text{dep}}} = r_x^{A_k} \lambda^{A_k} \\
		s_x^{A_{k,\text{dep}}} = s_x^{A_k} \lambda^{A_k}
		\end{matrix} \right . .
\end{equation}

With the mapping (\ref{eq:Mapping}), (\ref{eq:PartiallyDependendCBF}) can be mapped to binomial opinions, where two binomial opinions $\omega_x^{A}$ and $\omega_x^{B}$ then are accumulated by the operator
\begin{equation}
	\omega_x^{A} \, \tilde{\oplus} \, \omega_x^{B} = (\omega_x^{A,\text{dep}} \, \underline{\oplus} \,  \omega_x^{B,\text{dep}}) \oplus \omega_x^{A,\text{ind}} \oplus \omega_x^{B,\text{ind}} \, .
\end{equation}

In order to account for the fact that information on the reliability of an agent may age over time, a probability sensitive trust discounting step is introduced with the probability $P_{\text{sa}} = P( \{ \textit{trust still valid}\} ) \approx 1$. Note that the trust aging step refers to the \textit{source} trust rather than the data reliability. The source trust can be calculated as

\begin{equation}
	\tilde{\omega}_{x,\text{aged}}^{A_k} = \left \{ \begin{array}{ll}
	\mathbf{b}_{x,\text{aged}}^{A_k} &= P_{\text{sa}} \mathbf{b}_x^{A_k} \\
	u_{x,\text{aged}}^{A_k} &= 1 - P_{\text{sa}} \sum \limits_{x \in \mathcal{R}(\mathbb{X})}  \mathbf{b}_x^{A_k} \\
	\mathbf{a}_{x,\text{aged}}^{A_k} &= \mathbf{a}_{x}^{A_k} 
	\end{array}\right . \, .
\end{equation}

\subsection{Trust revision} \label{TrustRevision}

In case of inconsistency between incoming opinions, the users will report the incidence to the broker. The trust into the misbehaving agents then is revised. Note that the users and the broker except for the trust revision step itself basically run the same algorithm. Essentially, the broker just checks whether the reported blame is justified or not and then acts upon this.
The algorithm consists of the following steps:
\begin{enumerate}
	\item \textbf{Trust discounting:} The incoming opinions are discounted by the source reliability and the spacial and temporal aging of the information $\mathbf{T}_g^{d(\text{loc},\text{loc0})}$ and $\mathbf{T}_t^{t-t_0}$ . More details on the aging mechanism of the data can be found in \cite{Meuser2019}. This step can be formulated as
	\begin{subequations}
		\begin{align}
			\mathbf{P}_{dis} &= \mathbf{P}_{A_k,src} \cdot \mathbf{T}_g^{d(\text{loc},\text{loc0})} \mathbf{T}_t^{t-t_0} \cdot \mathbf{p}_0 \\
			\tilde{\omega}_{x,dis}^{A_k} &= \left \{ \begin{array}{ll}
			\mathbf{b}_{x,dis}^{A_k} &= \mathbf{P}_{dis} \mathbf{b}_x^{A_k} \\
			u_{x,dis}^{A_k} &= 1 - \mathbf{P}_{dis} \sum \limits_{x \in \mathcal{R}(\mathbb{X})}  \mathbf{b}_x^{A_k} \\
			\mathbf{a}_{x,dis}^{A_k} &= \mathbf{a}_{x}^{A_k} 
			\end{array}\right . \, .
		\end{align}
	\end{subequations} 
	\item \textbf{Clustering:} The incoming opinions are clustered with respect to their pairwise degree of conflict
	\begin{equation}
		\small
		\textit{DC}_x^{A_j,A_k} = \frac{1}{2} \sum \limits_{\forall j, k > j} | P_{x,dis}^{A_j} - P_{x,dis}^{A_k} | (1-u_{x,dis}^{A_j}) (1-u_{x,dis}^{A_k}) \, .
	\end{equation}
	Therefore, the set of triples $\{ (A_j, \, A_k, \, \textit{DC}_{A_j,A_k}) \}$ is sorted from small to big with respect to $\textit{DC}_x^{A_j,A_k}$ 
	 \begin{equation}
		\mathcal{L}_{\textit{DC}} = \left \{ (A_j, \, A_k, \, \textit{DC}_x^{A_j,A_k})_{l=1 \dots L}  \right\} \, ,
	 \end{equation}
	 and cut, when $\textit{DC}_x^{A_j,A_k} > \theta_{\text{DC}}$, where $\theta_{\text{DC}} \in [0,1]$ is a threshold parameter.
	 The triplets within $\mathcal{L}_{\textit{DC}}$ then form a (not necessarily connected) graph $\mathcal{G}(\mathcal{S}_A,\{ \textit{DC}_x^{A_j,A_k}\} )$, where the agents correspond to the vertices, the degrees of conflict between their opinions correspond to the edges, and each connected subgraph $\mathcal{SG}(\mathcal{G})$ represents a competitive hypothesis about the observation $x$. Hence, the subgraph $\mathcal{SG}^{\ast}(\mathcal{G})$ containing the largest number of agents stands for the hypothesis with the most support. Thus, all other subgraphs with less agents are pruned. Note, however, that  $\mathcal{SG}^{\ast}(\mathcal{G})$ is not necessarily unique. In this case, all $k$ equally supported hypotheses are considered.
	\item  \textbf{Calculation of the reference opinion:} If all vertices of $\mathcal{SG}^{\ast}(\mathcal{G})$ are connected with the same dominant vertex, the respective opinion is taken as reference opinion. Otherwise the reference opinion is calculated by
	\begin{equation}
		\omega_x^{\text{ref}_k} = \underset{\forall j \in \mathcal{SG}^{\ast}(\mathcal{G})} {\underline{\bigoplus}} \, \omega_{x,\text{aged}}^{A_j} \, .
	\end{equation}
	\item \textbf{Recalculation of degree of conflict:} In the next step, the degree of conflict is calculated between each available opinion and the reference opinion
	\begin{equation}
		\small
		\textit{DC}_x^{A_j,\text{ref}_k} = \frac{1}{2} \sum \limits_{\forall j} | P_{x,\text{aged}}^{A_j} - P_{x}^{\text{ref}_k} | (1-u_{x,\text{aged}}^{A_j}) (1-u_{x}^{\text{ref}_k}) \, \, .
	\end{equation}
	\item \textbf{Behaviour classification:} In order to classify whether an agent's opinion is valid or not, $\textit{DC}_x^{A_j,\text{ref}_k}$ is used as measure. As long as $\textit{DC}_x^{A_j,\text{ref}_k} < \theta_{DC}$, the opinion $\omega_{x,\text{aged}}^{A_j}$ is considered correct within the statistical variation and the respective agent is considered honest. In turn, if $\textit{DC}_x^{A_j,\text{ref}_k} > \theta_{\textit{DC}}$, the opinion significantly deviates from the reference opinion, which is considered the truth. Hence, the respective agent is classified as misbehaving. This leads to the partitioning of $\mathbb{A}$ into the subsets
	\begin{subequations}
		\begin{align}
			\mathcal{S}_{\text{honest}}^{\text{ref}_k} &= \left \{ A_j \in \mathbb{A} | \textit{DC}_x^{A_j,\text{ref}_k} \leq \theta_{DC}  \right \} \, ,\\
			\mathcal{S}_{\text{misb}}^{\text{ref}_k} &= \left \{ A_j \in \mathbb{A} | \textit{DC}_x^{A_j,\text{ref}_k} > \theta_{DC}  \right \} \, .		
		\end{align}
	\end{subequations}
	In the presence of multiple references, following the law of parsimony, the correct hypothesis is determined by
	\begin{equation}
		\text{ref}^{\ast} = \arg \max_{k} \bigg \{ \text{card} \{ \mathcal{S}_{\text{honest}}^{\text{ref}_k} \} \bigg \} \, .
	\end{equation} 
	\item \textbf{Trust revision:} While users only report other misbehaving agents, the broker revises the trust in agents classified as misbehaving. The simplest way of doing so would be to directly ban the agents classified as misbehaving from the system. However, if misbehavior can occur occasionally, this policy would be too harsh. Hence, the trust revision mechanism presented in \cite{Joesang2017} is used:
	\begin{subequations}
		\begin{align}
			\textit{MC} &= \max\limits_{\forall A_j \in \mathbb{A} } \left\{ (\textit{DC}_x^{A_j,ref^{\ast}} \right\} \, , \\
			\textit{AC} &= \frac{1}{\text{card} \{\mathbb{A}\}} \sum \limits_{\forall A_j \in \mathbb{A}} \textit{DC}_x^{A_j,ref^{\ast}} \, , \\				
			\textit{RW}_{A_k} &= \frac{\textit{MC} \cdot (\textit{DC}_x^{A_k,ref^{\ast}} - \textit{AC})}{\textit{MC} - \textit{AC}} \, , \\
			\tilde{\omega}_{A_k} &= \left \{ \begin{array}{l}
			\tilde{b}_{A_k} = (1- \textit{RW}_{A_k})b_{A_k} \\
			\tilde{d}_{A_k} = (1- \textit{RW}_{A_k})d_{A_k} \\
			\tilde{u}_{A_k} = (1- \textit{RW}_{A_k})u_{A_k} \\
			\end{array} \right . \, ,
		\end{align}
		\label{eq:TrustRevision}
	\end{subequations}	
\end{enumerate}
where $A_k \in \mathcal{S}_{\text{misb}}^{\text{ref}^{\ast}}$ are the agents classified as misbehaving and $\tilde{\omega}_{A_k}$ is the respective reviced trust into agent $A_k$. In (\ref{eq:TrustRevision}), $\textit{MC}$ stands for mean conflict, $\textit{AC}$ for average conflict, and $\textit{RW}$ for revision weight, respectively. 


\section{SIMULATIONS}

In this section, the proposed trust management and misbehavior detection mechanism is evaluated through simulation.
First the applicability of the mechanism to the exemplary intersection from the introduction is demonstrated in simulation, then a large scale simulation is given showing that the mechanism scales well with the size of the MAS.
For the intersection example, \cite{Mueller2019} describes the consecutive work with a focus on safty based on a real world example.

\subsection{Merging into an intersection with occlusion using cooperative information} \label{IntersectionExample}

\begin{figure}
	\centering
	\includegraphics[width=0.35\textwidth]{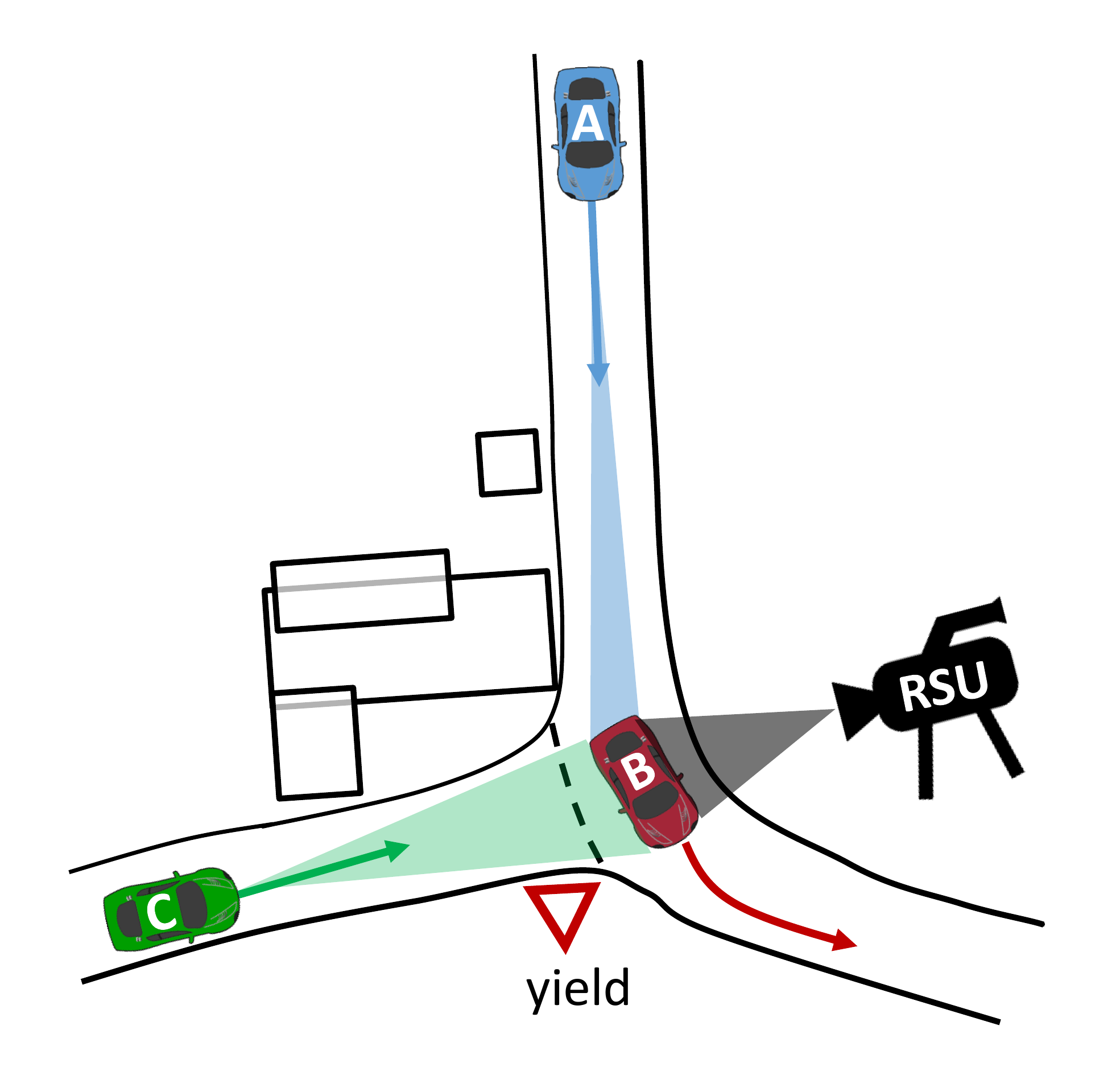}
	\caption{Intersection scenario.}
	\label{fig:Scenario}
\end{figure}

In order to demonstrate the effects of the misbehavior detection mechanism in detail, the mechanism is applied to the intersection scenario from the introduction and evaluated by simulation.
Consider the intersection scenario as depicted in Figure \ref{fig:Scenario}, where the vehicle C is approaching an intersection with occlusions, aiming to merge. Vehicle A follows vehicle B on the main road having the right of way. An RSU observes the whole intersection. All four agents measure the position $x$ of vehicle B.
For simplicity, in this paper, only the 1D position of vehicle B on its lane is considered and the motion is assumed to be compensated already by a feed-forward control. 
Hence, $x=0$ means that vehicle B is at its predicted position. For the measurement of B's position, Gaussian noise with a mean of $\mu = 0.25$ m and a standard deviation of $\sigma = 0.75$ m is assumed.

The agents build their opinions on $x$ using the following algorithm:
\begin{enumerate}
	\item Each agent transforms the measurements $X$ according to $Z = \frac{X - \hat{\mu}}{ \hat{\sigma} }$. Hence for the estimates $\hat{\mu} = \mu$ and $\hat{\sigma} = \sigma$, $Z \sim \mathcal{N}(z|0,1)$.
	\item Each agent estimates the distribution of $Z$ using a histogram. The histogram is mapped to the agent's opinion $\omega_x^A$ according to (\ref{eq:Mapping}) and $\omega_x^A$ is published.
\end{enumerate}
If inconsistency is reported, the broker performs the misbehavior detection mechanism presented in Section \ref{TrustRevision}. The RSU and the vehicles A and B create their opinion based on $N=50$ measurements, which corresponds to vehicle B been sensed for $5$ s at a typical sampling rate of $10$ Hz. Due to the occlusion, vehicle C only gets $N_C=10$ measurements, corresponding to vehicle C sensing vehicle B only for $1$ s. The following scenarios are evaluated:
\begin{enumerate}
	\item All agents behave correctly.
	\item The RSU is faulty. Hence $\hat{\sigma}_{RSU} \neq \sigma, \, \hat{\mu}_{RSU} \neq \mu$. The RSU detects its fault by applying the misbehavior detection mechanism and recalibrates itself, i.e. estimates $\hat{\mu}_{\text{RSU}}$ and $\hat{\sigma}_{\text{RSU}}$ again.
	\item The vehicles A and B launch a cooperative attack to discredit the RSU. Both of them use $\hat{\mu} = 1.0 \, \text{m}, \, \hat{\sigma} = 0.75 \, \text{m}$, unaware that vehicle C is approaching as well. A word-against-word situation builds up that is resolved by comparing the opinions about vehicle C's position $x_c$.
\end{enumerate}

\begin{figure}[thpb]
	\centering
	\begin{tikzpicture}
	\footnotesize
%
%

\begin{axis}[%
width=0.4 \textwidth,
height=3cm,
at={(0.0cm,0.0cm)},
scale only axis,
xmin=0,
xmax=0.3,
xlabel style={font=\color{white!15!black}},
xlabel={Threshold parameter},
ymin=0,
ymax=1,
y label style={at={(axis description cs:0.05,0.5)}},
xtick={0,0.05,0.1,0.15,0.2,0.25,0.3},
xticklabels = {$0$,$0.05$,$0.1$,$0.15$,$0.2$,$0.25$,$0.3$},
ylabel={Correct Classification Rate},
axis background/.style={fill=white},
axis x line*=bottom,
axis y line*=left,
xmajorgrids,
ymajorgrids,
legend style={at={(0.5,1.03)}, anchor=south, legend cell align=left, align=left, draw=white!15!black}
]
\addplot [color=blue, line width=2.0pt]
  table[row sep=crcr]{%
0.1	0.454\\
0.11	0.575\\
0.12	0.566\\
0.13	0.658\\
0.14	0.668\\
0.15	0.71\\
0.16	0.692\\
0.17	0.687\\
0.18	0.673\\
0.19	0.633\\
0.2	0.573\\
0.21	0.52\\
0.22	0.44\\
0.23	0.363\\
0.24	0.293\\
0.25	0.218\\
0.26	0.178\\
0.27	0.137\\
0.28	0.082\\
0.29	0.059\\
0.3	0.042\\
};
\addlegendentry{Collaborative Attack detected}

\addplot [color=blue, dashed, line width=2.0pt]
  table[row sep=crcr]{%
0.1	0.561\\
0.11	0.692\\
0.12	0.672\\
0.13	0.779\\
0.14	0.768\\
0.15	0.823\\
0.16	0.801\\
0.17	0.816\\
0.18	0.836\\
0.19	0.802\\
0.2	0.764\\
0.21	0.731\\
0.22	0.705\\
0.23	0.648\\
0.24	0.562\\
0.25	0.481\\
0.26	0.448\\
0.27	0.384\\
0.28	0.289\\
0.29	0.241\\
0.3	0.177\\
};
\addlegendentry{At least one attacker identified}

\addplot [color=red, dashdotted, line width=2.0pt]
  table[row sep=crcr]{%
0.1	0.436\\
0.11	0.304\\
0.12	0.315\\
0.13	0.196\\
0.14	0.193\\
0.15	0.117\\
0.16	0.117\\
0.17	0.098\\
0.18	0.062\\
0.19	0.059\\
0.2	0.057\\
0.21	0.059\\
0.22	0.038\\
0.23	0.034\\
0.24	0.046\\
0.25	0.041\\
0.26	0.025\\
0.27	0.018\\
0.28	0.025\\
0.29	0.022\\
0.3	0.016\\
};
\addlegendentry{Successfull wrong accusation}

\addplot [color=black, line width=2.0pt]
  table[row sep=crcr]{%
0	0\\
0.01	0\\
0.02	0\\
0.03	0\\
0.04	0.001\\
0.05	0.008\\
0.06	0.048\\
0.07	0.129\\
0.08	0.195\\
0.09	0.324\\
0.1	0.467\\
0.11	0.625\\
0.12	0.679\\
0.13	0.806\\
0.14	0.855\\
0.15	0.911\\
0.16	0.938\\
0.17	0.958\\
0.18	0.98\\
0.19	0.992\\
0.2	0.997\\
0.21	0.995\\
0.22	0.998\\
0.23	1\\
0.24	1\\
0.25	1\\
0.26	1\\
0.27	1\\
0.28	1\\
0.29	1\\
0.3	1\\
};
\addlegendentry{All Agent Honest}

\end{axis}
	\end{tikzpicture}
	\caption{Classification rate as function of threshold in the collaborative attack scenario.}
	\label{fig:Threshold}
\end{figure}
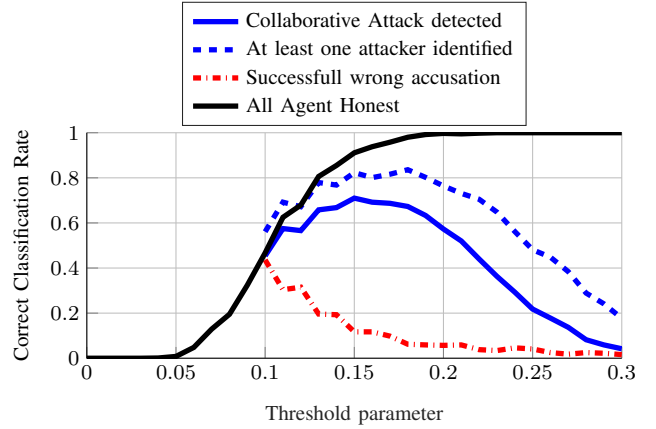

The results of scenario 1 and 3 are summarized within Figure \ref{fig:Threshold}.
It can be seen from Figure \ref{fig:Threshold} that $\theta_{\textit{DC}} \in [0.15,0.2]$ is a reasonable choice. For $\theta_{\textit{DC}} < 0.15$, the misclassification rate is too high as statistically insignificant deviations from the reference opinion are considered a thread, while for $\theta_{\textit{DC}} > 0.2$ the misbehavior detection mechanism gets insensitive. For $\theta_{\textit{DC}} \in [0.15,0.2]$, the detection probability even of a collaborative attack is at about $70 \%$, the probability of at least identifying one of the agents as misbehaving is at $80 \%$, while the probability that the collaborative attack is successful is between $5.7\%$ and $11.7\%$. Thus, the misbehavior detection mechanism is capable to cope with such collaborative attacks even for this worst case scenario.

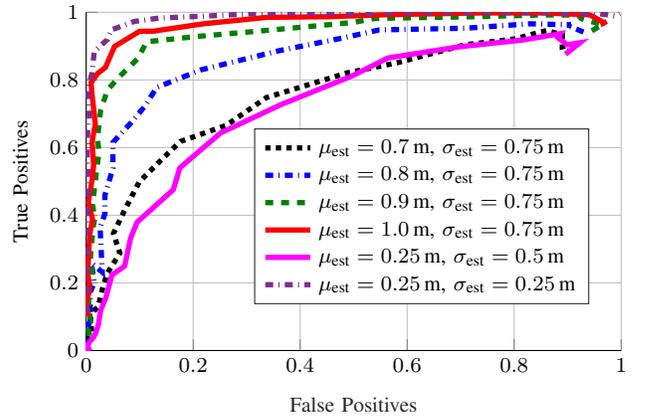
\begin{figure}[thpb]
	\centering
	\begin{tikzpicture}
	\footnotesize
	\definecolor{mycolor1}{rgb}{1.00000,0.00000,1.00000}%
	\definecolor{mycolor2}{rgb}{0.49000,0.18000,0.56000}%
%
%

%

\begin{axis}[%
width=0.4 \textwidth,
height=4.5cm,
at={(0.0cm,0.0cm)},
scale only axis,
xmin=0,
xmax=1,
xlabel style={font=\color{white!15!black}},
xlabel={False Positives},
ymin=0,
ymax=1,
y label style={at={(axis description cs:0.05,0.5)}},
ylabel={True Positives},
axis background/.style={fill=white},
axis x line*=bottom,
axis y line*=left,
xmajorgrids,
ymajorgrids,
legend style={at={(0.314,0.137)}, anchor=south west, legend cell align=left, align=left, draw=white!15!black}
]
\addplot [color=black, dotted, line width=2.0pt]
  table[row sep=crcr]{%
0.002	0.003\\
0.003	0.007\\
0.002	0.011\\
0.001	0.009\\
0.001	0.019\\
0.004	0.032\\
0.003	0.038\\
0.005	0.045\\
0.01	0.056\\
0.008	0.1\\
0.017	0.122\\
0.022	0.128\\
0.03	0.18\\
0.039	0.227\\
0.063	0.288\\
0.051	0.346\\
0.071	0.412\\
0.099	0.498\\
0.138	0.56\\
0.176	0.618\\
0.262	0.666\\
0.337	0.748\\
0.488	0.821\\
0.582	0.851\\
0.708	0.904\\
0.8	0.921\\
0.86	0.946\\
0.89	0.934\\
0.89	0.898\\
0.912	0.905\\
0.902	0.885\\
};
\addlegendentry{$\mu_{\text{est}}= 0.7 \, \text{m, }\sigma_{\text{est}}= 0.75 \, \text{m}$}

\addplot [color=blue, dashdotted, line width=2.0pt]
  table[row sep=crcr]{%
0.003	0.011\\
0.003	0.012\\
0.005	0.028\\
0.004	0.033\\
0.006	0.055\\
0.004	0.075\\
0.012	0.103\\
0.008	0.1\\
0.007	0.171\\
0.015	0.2\\
0.01	0.258\\
0.03	0.232\\
0.026	0.364\\
0.035	0.404\\
0.035	0.459\\
0.05	0.539\\
0.05	0.615\\
0.083	0.672\\
0.109	0.721\\
0.132	0.778\\
0.215	0.83\\
0.362	0.887\\
0.476	0.923\\
0.548	0.948\\
0.709	0.953\\
0.829	0.966\\
0.906	0.963\\
0.897	0.963\\
0.932	0.94\\
0.933	0.932\\
0.935	0.934\\
};
\addlegendentry{$\mu_{\text{est}}= 0.8 \, \text{m, } \sigma_{\text{est}}= 0.75 \, \text{m}$}

\addplot [color=black!50!green, dashed, line width=2.0pt]
  table[row sep=crcr]{%
0.006	0.043\\
0.004	0.063\\
0.005	0.086\\
0.008	0.087\\
0.005	0.133\\
0.003	0.181\\
0.008	0.205\\
0.011	0.231\\
0.008	0.276\\
0.013	0.365\\
0.016	0.39\\
0.012	0.431\\
0.017	0.501\\
0.024	0.588\\
0.021	0.626\\
0.027	0.71\\
0.04	0.771\\
0.065	0.814\\
0.097	0.864\\
0.113	0.913\\
0.21	0.929\\
0.334	0.946\\
0.499	0.97\\
0.54	0.982\\
0.73	0.987\\
0.849	0.992\\
0.919	0.992\\
0.926	0.981\\
0.959	0.966\\
0.965	0.964\\
0.945	0.945\\
};
\addlegendentry{$\mu_{\text{est}}= 0.9 \, \text{m, } \sigma_{\text{est}}= 0.75 \, \text{m}$}

\addplot [color=red, line width=2.0pt]
  table[row sep=crcr]{%
0.004	0.101\\
0	0.146\\
0.006	0.174\\
0.006	0.185\\
0.003	0.235\\
0.006	0.325\\
0.007	0.347\\
0.012	0.387\\
0.004	0.438\\
0.012	0.514\\
0.014	0.551\\
0.011	0.615\\
0.017	0.671\\
0.011	0.757\\
0.009	0.792\\
0.022	0.82\\
0.036	0.837\\
0.053	0.899\\
0.1	0.944\\
0.125	0.944\\
0.219	0.966\\
0.337	0.985\\
0.457	0.987\\
0.566	0.993\\
0.732	0.998\\
0.854	0.998\\
0.936	0.993\\
0.942	0.992\\
0.965	0.972\\
0.966	0.967\\
0.973	0.973\\
};
\addlegendentry{$\mu_{\text{est}} = 1.0 \, \text{m, } \sigma_{\text{est}} = 0.75 \, \text{m}$}

\addplot [color=mycolor1, line width=2.0pt]
  table[row sep=crcr]{%
0.001	0.001\\
0	0.002\\
0.001	0\\
0.002	0.002\\
0.003	0.003\\
0.005	0.005\\
0.003	0.011\\
0.003	0.018\\
0.005	0.023\\
0.013	0.038\\
0.018	0.051\\
0.023	0.078\\
0.027	0.118\\
0.037	0.157\\
0.048	0.223\\
0.072	0.25\\
0.083	0.332\\
0.095	0.379\\
0.163	0.476\\
0.175	0.54\\
0.251	0.643\\
0.364	0.727\\
0.496	0.811\\
0.564	0.865\\
0.697	0.898\\
0.816	0.918\\
0.887	0.937\\
0.882	0.925\\
0.899	0.904\\
0.923	0.91\\
0.896	0.876\\
};
\addlegendentry{$\mu_{\text{est}}= 0.25 \, \text{m, } \sigma_{\text{est}} = 0.5 \, \text{m}$}

\addplot [color=mycolor2, dashdotted, line width=2.0pt]
  table[row sep=crcr]{%
0	0.109\\
0.001	0.139\\
0.002	0.16\\
0.001	0.199\\
0.001	0.251\\
0.002	0.272\\
0.001	0.357\\
0.001	0.373\\
0.004	0.453\\
0.003	0.534\\
0.002	0.58\\
0.003	0.671\\
0.006	0.685\\
0.003	0.772\\
0.011	0.821\\
0.015	0.884\\
0.04	0.92\\
0.047	0.948\\
0.09	0.973\\
0.136	0.983\\
0.212	0.988\\
0.342	0.998\\
0.517	0.996\\
0.581	0.999\\
0.734	0.999\\
0.867	0.999\\
0.951	0.994\\
0.959	0.996\\
0.985	0.99\\
0.988	0.989\\
0.997	0.997\\
};
\addlegendentry{$\mu_{\text{est}} = 0.25 \, \text{m, } \sigma_{\text{est}} = 0.25 \, \text{m}$}

\end{axis}
	\end{tikzpicture}
	\caption{ROC curve for the misbehaving RSU scenario (scenario 2).}
	\label{fig:ROC}
\end{figure}

Figure \ref{fig:ROC} shows the Receiver Operation Characteristics (ROC) curve for different estimations of $\hat{\mu}$ and $\hat{\sigma}$. The closer an ROC curve gets to $(0,1)$, the better. It shows that even estimation errors smaller than the standard deviation can be efficiently detected at low false positive rate.

When the RSU is classified as misbehaving, it recalibrates itself based upon its wrong opinion and the reference opinion, according to
\begin{subequations}
	\begin{align}
		\hat{\mu} &= \overline{\mathbf{z}}^{\top} \mathbf{P_{\text{RSU}}} - \overline{\mathbf{z}}^{\top} \mathbf{P_{\text{ref}}} \, ,\\
		\hat{\sigma} &=  \sqrt{ \sum \limits_{\forall i} ( \, (\overline{z}_i - \hat{\mu})P_{\text{RSU},i} \, ) ^2 }	\, ,
	\end{align}	
\end{subequations}
where $\overline{\mathbf{z}}$ are the centers of the histogram bins. The recalibration has been evaluated over $1000$ runs for $\theta_{\textit{DC}} = 0.15$. In average, $\hat{\mu} = 0.250$ m and $\hat{\sigma} = 0.765$ m resulted. Thus the mechanism not only detects the fault but corrects it as well.

\subsection{Traffic jam detection}

\begin{figure} [htb]
	\centering
	\includegraphics[width=0.3\textwidth]{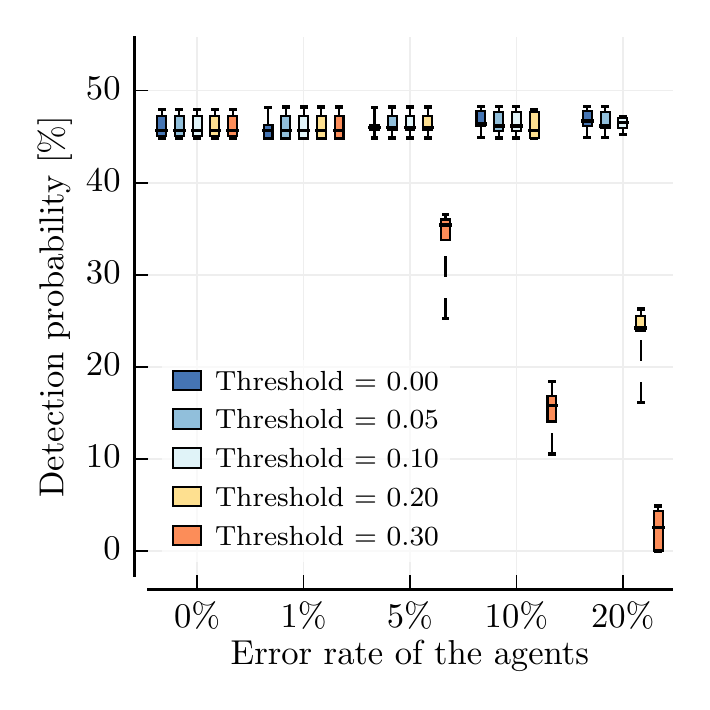} \\
	\includegraphics[width=0.3\textwidth]{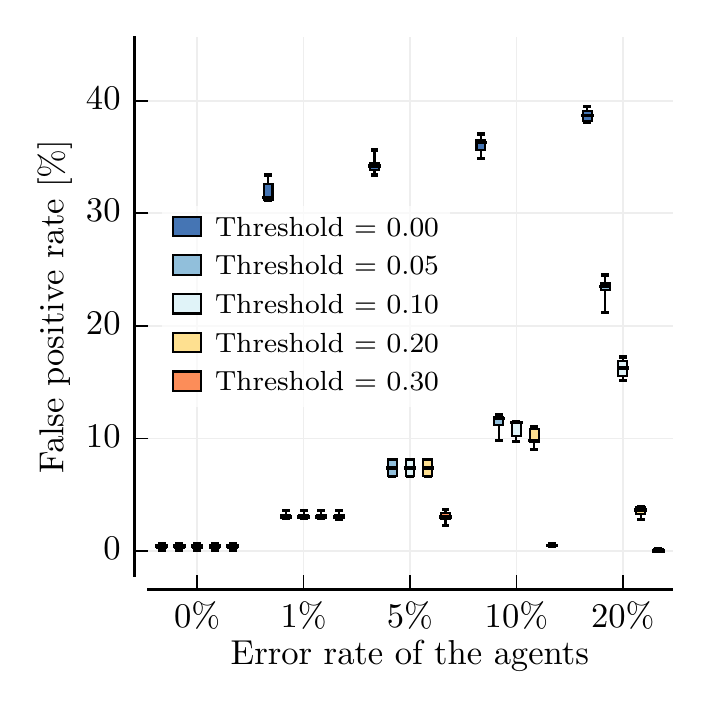}
	\caption{Detection probability and False positive rate in dependence of the threshold $\theta_{\textit{DC}}$ and the error rate of the agents resulting from the large scale traffic simulation of Cologne. }
	\label{fig:LargeScale}
\end{figure}

In order to demonstrate that the presented approach scales well with the size of the MAS and to gain high statistical certainty, the presented algorithm is evaluated on a large scale scenario simulating the whole traffic of Cologne. As simulation environment, Simonstrator \cite{Meuser2019B} and SUMO \cite{Krajzewicz2012} are used. In the scenario, $10\%$ of the agents are assumed to be misbehaving.
Figure \ref{fig:LargeScale} shows the detection probability $p_{\text{tp}}$ as well as the false positive rate $p_{\text{fp}}$ within each measurement in dependence on the agents' error rate and the threshold $\theta_{\textit{DC}}$. The results describe the agents' reporting behavior, hence the presented algorithm is run without the trust revision step.
For example, it can be seen from Figure \ref{fig:LargeScale} that if $10\%$ of the data reported by a correctly behaving agent is wrong and $\theta_{\textit{DC}}=0.1$, the  detection probability is about $p_{\text{tp}}\approx45 \%$, while $p_{\text{fp}}\approx 10\%$.
For a rough illustrative calculation, it is assumed that the trust into an agent is depleted if three misbehaviors are detected within a batch of three consecutive reports, while the trust is restored for each batch. Then, the probabilities of wrongly blaming a correctly behaving agent $p_{\text{wb}}$ and the probability of detecting a misbehaving agent $p_{\text{dm}}$ are given by
\begin{equation}
p_{wb} = 1 - (1 - p_{\text{fp}}^3)^n \quad \text{and} \quad p_{\text{dm}} = 1-(1 - p_{\text{tp}}^3)^n \, ,
\end{equation}
respectively, where n is the number of batches.
Hence, after $n=15$, i.e. after a total of $45$ reports, $p_{wb}=76.1\%$ of the misbehaving agents are eliminated from the system, while only $1.49\%$ of honest agents are excluded. 
At typical measurement update rates of $10$ Hz, even if several measurements are bundled to a single reported opinion and even if there are time gaps between the reports, it is a matter of minutes till the vast majority of misbehaving agents is excluded from the system.
This illustrates that the misbehavior detection mechanism works well. 

\section{CONCLUSIONS}

In this work, an trust management and misbehavior detection mechanism has been presented that, based on subjective logic, amends reliability information to the cooperative information spread through the system. The mechanism is resource efficient and robust against faulty and even intentionally malicious agents. It has been shown that the mechanism is abled to quickly detect and remove such misbehaving agents from the system and thus a high overall reliability can be provided.

In future work, the presented mechanism might be transfered to other network topologies.

\addtolength{\textheight}{-12cm}   

%

%


\bibliographystyle{IEEEtran}
\bibliography{ICCA2019}

\end{document}